# Solid solutions with coexisting antiferroelectric and ferroelectric phases as materials with negative refractive index

V.M. Ishchuk and V. L. Sobolev[*1]

Science and Technology Center "Reaktivelektron", of the National Academy of Sciences of Ukraine, 83096 Donetsk, Ukraine
[*]Department of Physics, South Dakota School of Mines and Technology, Rapid City, SD 57701.

## Abstract

A possibility of use of the controlled decomposition of solid solutions of oxides with perovskite structure in the state of coexisting domains of the antiferroelectric (AFE) and ferroelectric (FE) phases for manufacturing of materials with negative refractive index is demonstrated. The lead zirconate titanate based solid solutions are considered as an example of substances suitable for creation of such materials. Manufactured composites constitute a dielectric matrix with a structure of conducting interphase boundaries separating domains of the FE and AFE phases. The electric conductivity of interphase boundaries occurs as a result of the local decomposition of solid solutions in the vicinity of these boundaries. The process of local decomposition and consequently the interphase boundaries conductivity can be controlled by means of external influences.

Keywords: refraction index; solid solutions; coexisting ferroelectric and antiferroelectric states.

---

[1] Correspondence: vladimir.sobolev@sdsmt.edu

The possibility of existence of materials with negative refractive index for electromagnetic radiation and the analysis of the core set of their physical properties was first reported in [1]. Materials with negative refractive index have been hypothetical for a long period until the first experimental confirmations of hypothesis expressed in Ref.1 appeared [2, 3]. Since that time researches have taken an active interest in materials of this kind and the number of publications on this topic constantly increases. Comprehensive reviews [4-7] of different physical properties as well as applications of negative refractive index materials have touched practically all aspects of this interesting part of material science existing up to date.

First materials possessing the negative refractive index only in high frequency region of electromagnetic radiation spectrum represented a composite construction of complex-shape elements (for example, a periodic array of split-ring resonators with wires placed uniformly between the split rings and so on) [2,3,8]. It is practically impossible to manufacture similar design for the optical range of spectrum.

The metamaterials with the negative refractive index in the near infrared and optical ranges of spectrum were the metal-dielectric composites [9-11]. The technological approach to manufacturing of composites mentioned in [9-11] is quite laborious and is difficult to reproduce. It has to be noted that the technological process for manufacturing of these composites stays the same even now.

We have developed a simple and reproducible method for fabrication of composite structures with periodic arrangement of high conductivity inclusions in dielectric matrix. These inclusions have sizes in the range of 8 to 12 *nm* with period of arrangement of 0.2 – 2.0 *μm*. The method is based on the process of decomposition of the solid solution in the vicinity of the interphase boundaries between domains of the coexisting ferroelectric (FE) and antiferroelectric (AFE) phases [12-14].

Lead zirconate titanate $Pb(Zr_{1-x}Ti_x)O_3$ (PZT) based solid solutions are the most studied substances in which the FE and AFE ordering takes place. These materials are characterized by a small difference in free energies of the FE and AFE states in the solid solution at specified concentrations of components ($PbZrO_3$ and $PbTiO_3$). Domains of the above-mentioned phases coexist in the sample volume of solid solutions belonging to this interval of specific compositions [15-18]. The metastable phase domains have the shape of cylindrical domains imbedded into the stable phase matrix [15, 16] in the thin crystals or the shape of ellipsoids of revolution close to spheres in the bulk crystals [19]. Such structure of coexisting domains is realized in the crystallites that are not subjected to any external strains. Under the lateral strains the cylindrical domain structure is transformed into the stripe-domain one.

The relative stability of the FE and AFE phases depends on the position of the solid solution in the diagram of its phase states (on the relation between the concentrations of components of the solid solution). One can change the relative stability of phases and along with it the volume share of phases in the sample changing the position of solid solution in this diagram. The sizes of the metastable phase domains and their density (the period of the domain structure) change when the share of this phase in the solid solution varies. The density and sizes of the domain structure can be also changed by means of variation of an external electric field intensity as well as mechanical stress (pressure, uniaxial or biaxial compressions or tensile stress). These external influences also change the relative stability of the FE and AFE phases.

Conductivity of the majority of oxides with perovskite crystal structure (the PZT-based solid solutions are among them) can be varied by means of ion substitutions in lattice sites in a very wide range from pure dielectric state (with resistivities of the order of $10^{14} - 10^{16}$ $\Omega \cdot cm$) to the conducting state (with resistivities of the order of 100 $m\Omega \cdot cm$).

Interphase boundaries (IPB) that separate the adjacent FE and AFE are characterized by continuous conjugation of crystal planes [15, 16, 20] (without discontinuities of crystal planes and dislocations at the interphase boundaries). Such coherent character has to be accompanied by an increase of elastic energy of the crystal lattice along these boundaries.

The equivalent positions of the crystal lattice in PZT-based solid solutions are occupied by ions with different sizes and/or different electric charges. In the bulk of each domain (away from the domain boundaries) the net force acting on each of these ions is zero. In the vicinity of the IPB the balance of forces is disturbed, and as a result the "large" ions are driven out into the domains with the larger configuration volume and correspondingly with the larger distances between atomic planes. At the same time the "small" ions are driven out into the domains with smaller distances between atomic planes. Such processes are accompanied the competition of a reduction in the elastic energy along the IPB on the one hand, and an increase of the energy caused by the deviation of the solid solution composition from the equilibrium composition on the other hand. The processes described above will be finished when the structure of the new IPBs will correspond to the minimum of energy. This new "dressed" IPBs are not a bare IPBs anymore. In what follows we will use the term IPB for the "dressed" interphase boundary which appears when the redistribution of ions already took place.

The *A*- and *B*-positions of the perovskite crystal lattice of the solid solutions are occupied by ions with different ionic charges ($Pb^{2+}$, $La^{3+}$, $Li^+$ in *A*-sites and $Zr^{4+}$, $Ti^{4+}$, $Nb^{5+}$, $Mg^{2+}$ in *B*-sites). Because of this the local decomposition of the solid solution along the IPBs can be accompanied by the local disturbance of electro-neutrality. Thus, the formation of the heterophase structure of coexisting domains is accompanied by a violation of the chemical homogeneity of the samples. The samples remain homogeneous at temperatures higher then $T_C$, when dipole-ordered phases are absent. The duration of the decomposition (establishment of the equilibrium heterogeneous composite structure) has the range from several hours to several tens of hours for different solid solutions [13, 14]. That is why one can effectively control the process of formation of the composite structure.

Segregates precipitated along the interphase boundaries are also PZT-base solid solutions [15] but their chemical composition is slightly different from the maternal solid solution. We select the chemical composition of the maternal solid solution and carry out the decomposition in such a way that segregates can possess a diverse set of physical properties (by means of control of the chemical composition of segregates). These segregates can be magnetic, dielectric or conducting.

Powder samples of the PZT-based solid solutions were manufactured by the co-precipitation of the components from the mixture of aqueous solutions of lead and lanthanum nitrates and chlorides of titanium and zirconium. After washing and drying the precipitates were annealed at $550°C$ and $850°C$.

Samples for optical studies were manufactured by hot pressing method (at a pressure of 30 MPa) at $1250°C$ during 8 hours. Lamellae with the thickness of 0.3 *cm* were cut from the sintered bars. These lamellae were grinded and polished. Before polishing the grinded lamellas were annealed at $1200°C$ in the presence of $PbZrO_3$ filling during one hour and after that at $1100°C$ in oxygen enriched atmosphere during one hour. After the polishing the lamellae were subjected to the second annealing at $850°C$ in oxygen enriched atmosphere during two hours. The wavelength dependences of the light transmission coefficient were measure using Hitachi U-4000 Spectrophotometer.

Initial PZT-based solid solutions have been manufactured with the resistivity of the order of $10^{14}$ $\Omega \cdot cm$ and compositions corresponding to the state of the coexisting domains of the FE and AFE phases in the samples' volume. Decomposition of solid solution was performed in such a way that segregates in the vicinity of IPBs had high (close to metallic) conductivity.

The dependences of transparency of one of such materials as a function of the wave length are presented in the figure below as an example.

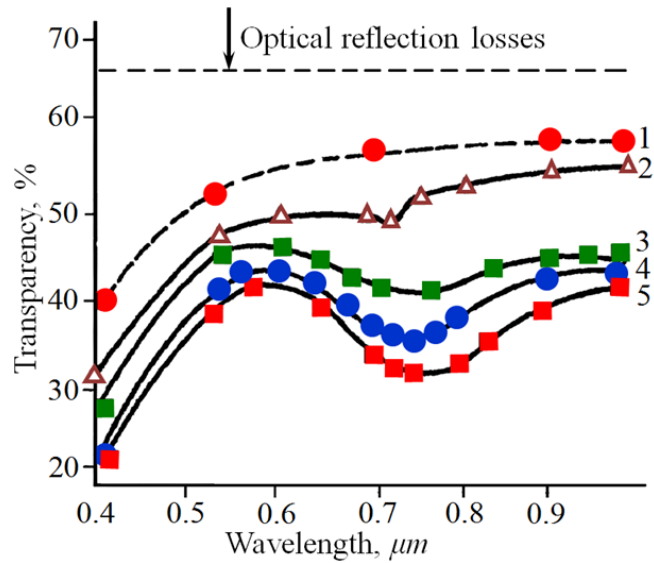

Dependence of the light transmission coefficient on the wave length for the transparent composite material (the dielectric matrix containing conductive metallic segregates). The increasing numbers near curves correspond to materials with the increasing value of conductivity of segregates.

The dips in the curves correspond to the presence of the negative refraction regime. The modulation depth can be controlled by changing the conductivity of segregates. It is also possible to select the optic wave length range by changing the position of the solid solution in the "composition-temperature" phase diagram of the substance. Another possibility is by changing the position of the solid solution in the "electric field-temperature" phase diagram by means of varying the potential difference between the element's electrodes.

However the most interesting results were obtained using thin film structures. The change of the region of negative refraction in this case can be achieved both by the application of the electric field to the film substrate, if the ferroelectric crystal is chosen as a substrate material, and by the flexural strains of the substrate. The modulation of transmitted light has been observed in both these situations.

Based on our results on use of the controlled decomposition of solid solutions for manufacturing of composite materials one can relatively easy develop substances and device components with controlled optical characteristics.